# An Integrative Introduction to Human Augmentation Science

Bradly Alicea

Orthogonal Research Lab (http://orthogonal-research.weebly.com); OpenWorm Foundation (http://openworm.org)

## ABSTRACT

Human Augmentation (HA) spans several technical fields and methodological approaches, including Experimental Psychology, Human-Computer Interaction, Psychophysiology, and Artificial Intelligence. Augmentation involves various strategies for optimizing and controlling cognitive states, which requires an understanding of biological plasticity, dynamic cognitive processes, and models of adaptive systems. As an instructive lesson, we will explore a few HA-related concepts and outstanding issues. Next, we focus on inducing and controlling HA using experimental methods by introducing three techniques for HA implementation: learning augmentation, augmentation using physical media, and extended phenotype modeling. To conclude, we will review integrative approaches to augmentation, which transcend specific functions.

## Introduction

For much of human history, augmenting the human body has been a theme of both fiction and technological achievement. The traditional view of augmentation has been as a functional extension of the human body mediated through a biomimetic physical medium (Barfield and Williams, 2017). Contemporary technological innovations such as the brain-computer or brain-machine interface (BCIs/BMIs – see Wolpaw, 2002; Wolpaw, 2012; Hochberg et.al, 2006; O'Doherty et.al, 2011; Nicolelis, 2011; Vieria et.al, 2013; Rouse & Schieber, 2015; Sexton, 2015), high-resolution virtual environments (VEs), and optimization algorithms have also made the notion of HA technologically explicit. Less understood are the cognitive and biological processes underlying embodied cybernetic systems capable of augmenting the human body, brain, and mind. To remedy this situation, I will advance the argument that embodied cybernetic systems are symbiotic technological-biological relationships (Licklider, 1960) that provide the basis for measurement and manipulation of human augmentation (HA). These systems are influenced by cognitive and life-history (biological) processes such as attentional capacity, expertise, aging, and generalized plasticity. Augmentation presents us with a range of cognitive and anatomical effects that can dynamically shape this symbiotic relationship, particularly over time (Moore, 2008).

This paper will proceed by discussing the historical roots of augmentation and the conceptual heritage that inform modern views. We will then discuss the concept of mitigation strategies (Regli et.al, 2005; Fuchs et.al, 2007) and their role in the efficacy of augmentation. Following from this is discussion of whether methods used to induce augmentation have a solid cognitive and biological basis. This leads us to explore three general approaches to augmentation: performance landscapes, augmentation using physical media, and the extended phenotype model. To conclude the article, a discussion of multilevel augmentation will provide the reader with an appreciation for how augmentation proceeds in a complex system. This might also lead to methods that characterize the longer-term effects of augmentation at multiple levels of biological complexity.

**Multiple Perspectives**

Augmentation has a rich intellectual history, and features both philosophical and empirical contributions from a wide variety of individuals. Some of the earliest HA concepts were Leonardo DaVinci's "Ornithopter" and "Vitruvian Man". W.R. Ashby and Jose Delgado (Figure 1) provide a more modern perspective on augmentation focused on control mechanisms serving as generalized mitigation strategies. Ashby's work on cybernetics and regulation has provided us with a machine called the homeostat. The homeostat is a precursor of modern Artificial Intelligence (AI) and Genetic Algorithms (GAs). Ashby demonstrated homeostatic (or self-) regulation in the form of an automata that could adapt to its environment (Ashby, 1952). An example of this implementation is shown in Figure 1 (LEFT). While somewhat successful as a demonstration of ultrastability in an interconnected system (homeostasis, see Battle, 2014), the homeostat was difficult to scale to large, complex systems (Cariani, 2009).

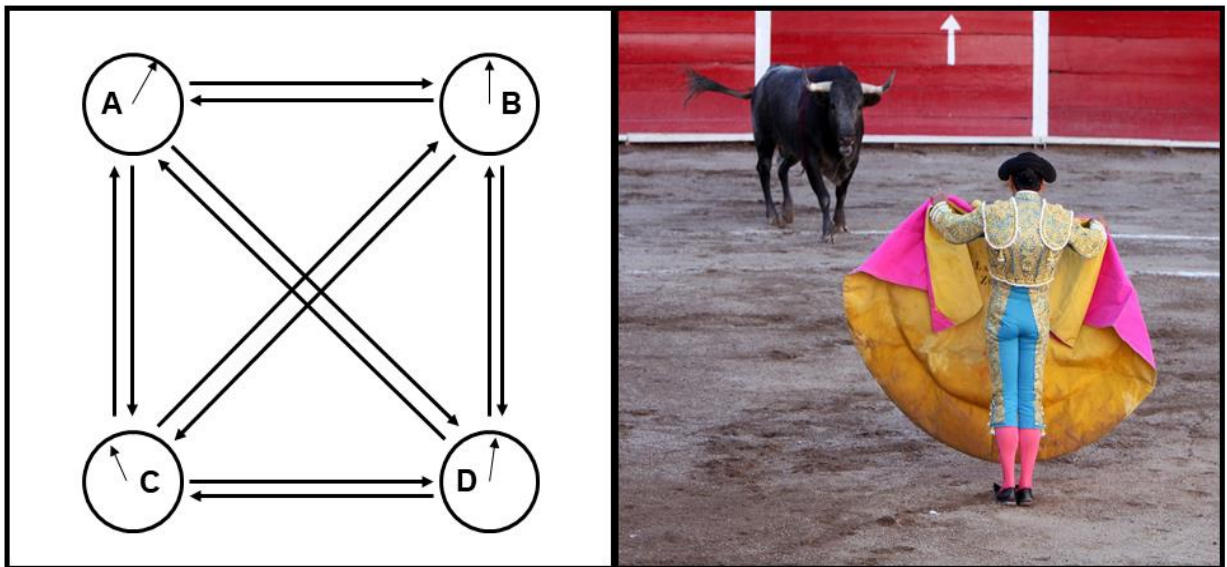

Figure 1. LEFT: an implementation of Ashby's homeostat. A-D are nominally independent regulatory systems (or governors) fully connected to all other systems. RIGHT: an image of a bullfight. When Delgado's hypothalamic switch is activated, the bull stops in its tracks and fails to charge the matador (picture from Wikimedia Commons).

While Ashby focused on artificial adaptive systems, Jose Delgado (1970) focused on modifying the brain itself to augment behavior. Delgado demonstrated this through bullfighting: a bull received an electrical impulse switch implant and then behaviorally stimulated by a bullfighter (see Figure 1, RIGHT). As the experimenter flipped a radio-controlled switch, the bull successfully stopped mid-charge (Marzullo, 2017). While ethically questionable, this and similar experiments by Delgado's contemporaries (Horgan, 2017) demonstrates how the simple delivery of electrical impulses can serve as a first-order control mechanism in a behavioral process.

**A More Practical Cutting-edge.** A later but perhaps more practical part of augmentation history comes from DARPA's Augmented Cognition project (Schmorrow and Kruse, 2002). As a legacy of Douglas Englebart's work in the 1960's (Englebart, 1962) and "Decade of the Brain" initiatives in the 1990s, Augmented Cognition (AC – Schmorrow and Stanney, 2009)) became an integral part of the Human-Computer Interaction (HCI) field. In particular, the development of HCI interfaces utilizing real-time brain measurements, cognitive state assessments, and a formalized closed-loop model has opened up new vistas in terms of our understanding of both human performance and brain function (Stanney et.al, 2009). Closed-loop systems based on real-time measurements of neural systems have also enabled adaptive approaches to task allocation using computational systems (Prinzel et.al, 2000). Interactions between neural systems and task environments enable the open-ended optimization of performance (Scallen and Hancock, 2001; Miller and Parasuraman, 2007). Since that time, we have witnessed consumer applications that create augmentative conditions, such as video games controlled by brain wave measurements and the development of wearable and even implantable sensors to measure cognitive/behavioral state.

**New Directions in Augmentation.** In parallel to the Augmented Cognition research program, there exist a number of innovative approaches to problems in Augmentation. Biomechatronic devices such as exoskeletons and orthoses augment the ability to perform physical tasks lost due to injury (Humphries, 2014; Herr, 2009). Various research groups have explored the use of BCIs and BMIs to reestablish direct communication between the brain and the outside world. In this case, communication with the outside world has involved the control of VEs using electroencephalogram (EEG) measurements (Wolpaw, 2002; Machado et.al, 2010), control of limb movements in VEs (Ifft et.al, 2013), prosthetic devices such as robotic arms (Hochberg et.al, 2006; O'Doherty et.al, 2011), wheelchairs (Rajamgan et.al, 2016), and even stimulation of other brains (Nicolelis, 2011; Vieira et.al, 2013). This has also involved work in animal models such as macaques and rodents to establish the role of brain stimulation and neural decoding in the production and control of behavioral states (Nicolelis, 2011). As in the case of the biomechatronic forms of augmentation, computers and brain implants serve to replace nerves and even muscle function lost to injury. More recent advances has enabled the sharing of neural signals between brains (O'Doherty et.al, 2011; Vieira et.al, 2013), enabling dyadic communication in a mode other than speech or nonverbal cues.

## How Augmentation Works

To be effective, we must employ augmentation in a strategic manner with respect to the underlying mechanisms and processes of human perception and biology. The reason for this is clear: augmentation is simply not effective unless it complements (rather than counters) existing processes. This forces us to make several design decisions, and therefore go beyond the relationship between human and machine into the realm of Human Factors Engineering (HFE). When complementary forms of augmentation are coupled with an adaptive component such as Artificial Intelligence (AI), it provides us with what is called a *mitigation strategy*. A mitigation strategy should contain two elements: it must be quantitative, and it must be robust to both noise and dynamic range. In terms of a quantitative description, a performance curve or set of equations will suffice. Agent-based and other generative approaches (Alicea, 2008a) may be acceptable, but come with their own caveats. The requirements of noise tolerance and dynamic range considerations mean that the phenomenon in question must be well-characterized.

## Yerkes-Dodson Model of Mitigation

A simple example of a mitigation strategy in action is an application of the Yerkes-Dodson (Y-D) law (Cohen, 2018) to mitigate attentional fluctuations during a dynamic task. The Y-D law provides us with a bivariate curve that measures the performance level given a specific amount of arousal. Y-D curves from an inverse "U" (Figure 2), with low levels of performance for both low and high degrees of arousal. This curve also presents us with a theoretical model in the sense that inverse "U" describes and other arousal in a way that is useful for characterizing continuous behavior. Yet more importantly, the convex portion of the "U" is an optimal point that assists in managing an individual's attentional resources. Arousal either rises or falls as a function of losing attentional focus due to fatigue or distraction. Mitigating this variability involves attenuating natural fluctuations in the underlying physiological system, and there are higher-dimensional systems that have a more complicated relationship with the perceptual world.

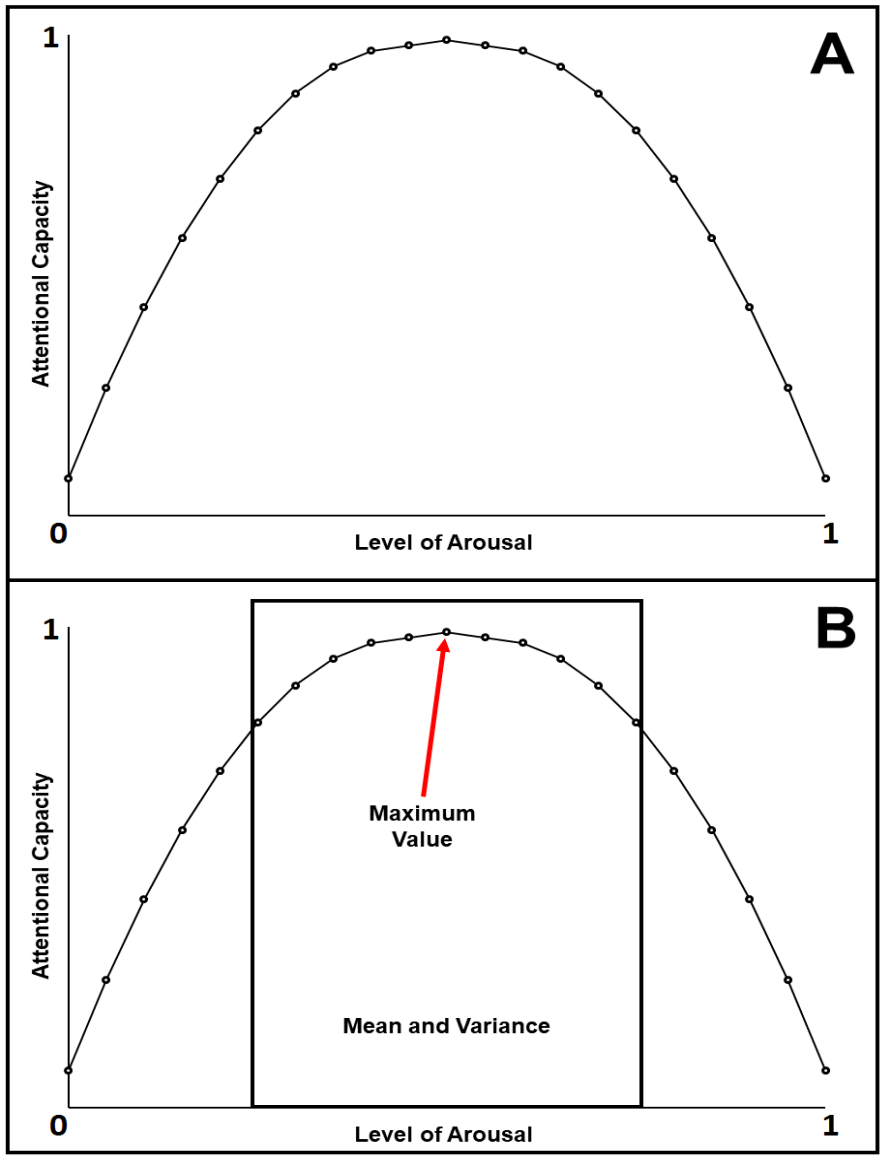

Figure 2. Yerkes-Dodson Law as a convex function and mitigation strategy. A) shape of a standard inverse U curve, B) maximum value for inverse U curve and bounding box for variance around maximum value centered upon the mean (the "mean-max" strategy).

**Mitigation as an Adaptive Mechanism.** The actual mitigation strategy itself involves creating adaptive heuristics for augmenting the attentional environment based on the dynamic range of observed behavioral outputs for a particular set of conditions. Returning to our inverse "U", we can create rules that impose augmentation under certain conditions. Mitigation itself can take many forms: from graphics on a display screen to auditory signals that refocus the individual to the task and thus changes the arousal. Optimizing the amount of arousal within a range of values follows a "mean-max" strategy. Using a heuristic in this way ultimately allows for instances of poor performance to be eliminated from the individual's repertoire. These instances often occur during the "dips and leaps" typical during skill-building (Gray & Lindstedt, 2017). When applied properly, mitigation strategies might even be able to mimic dynamic adaptive heuristics such as the gaze heuristic often employed by baseball players during outfield practice (Gigerenzer & Gray, 2017). Such heuristics allow for the details of the task subsumed by procedural memory and allow for the task itself to become intuitive.

**Generalized Adaptive and Closed-loop Mitigation.** When inducing augmentation with physical (or digital) media, one type of analysis we can use involves techniques inspired by linear control theory and adaptive computing. As a means of enforcing mitigation in a Yerkes-Dodson cognitive regime, linear state-space modeling is the most straightforward way to provide functional feedback in a closed-loop interface (Rouse and Schieber, 2015). In fact, this is the predominant type of mitigation strategy used in the design of brain-computer interfaces (Figure 2). Kaber et.al (2010) discuss this in terms of human-centered automation theory, which argues that performance is constrained by cognitive load. Cognitive load is a tradeoff between the number of tasks and the level of automation, and so performance is optimized by minimizing the number of tasks presented to the individual at any one time.

## Intelligence Augmentation

Another instance of HA involves various symbiotic human interactions with AI. Engelbart (1962) used the term intelligence augmentation (IA) to characterize these interactions. Rather than simply being the inverse of AI, IA provides a way to complement AI systems in ways that complement the inherent weaknesses of both human and machine intelligence. Opportunities for novel research exists in bringing together both AI and IA, as AI and HCI (an allied field of IA) have been divorced from one another during the past 25 years (Grudin, 2009). Even in the case of so-called strong AI, machine intelligence may not be able to emulate or even represent certain dimensions of human intelligence (Braga and Logan, 2017).

**Augmentation and Multidimensional Learning and Memory**

One such underappreciated model of augmentation involves learning, memory, and multisensory integration. These cognitive domains are much different in terms of their function and effect on behavior, and characterized through a rigorous, quasi-quantitative model of optimal function. Stefanucci et.al (2007) present an experimental setting called the InfoCockpit. The InfoCockpit is a flight simulator that presents visual, auditory, and proprioceptive information in a spatially explicit manner, and sits in relation to the head and upper body. Congruent information, or multisensory cues presented in the same spatial location (Stein & Meredith, 1993), has been found to improve recall by 131% (Stefanucci et.al, 2007). This is due to a phenomenon called multisensory integration (Stein & Meredith, 1993; Lewkowicz & Ghazanfar, 2009) that operates in a manner similar to the Y-D model of attentional mitigation.

One benefit for using spatially explicit multisensory models to mitigate learning and memory performance is that we can use both spatial/temporal congruence and the principle of inverse effectiveness to enhance recall. When cues are delivered in two different sensory modalities (e.g.

visual and auditory), they can interfere with one another unless delivered at the same time and location in space. Congruence serves to integrate cues in an environmental context, and produces a synthetic effect that is greater than the sum of each sensory cue being presented alone. According to the principle of inverse effectiveness, we can enhance this superadditive effect in cases where single sensory cues are relatively weaker, so that the synthetic effect is stronger as the singular effect gets weaker or is masked by noise (Stein & Meredith, 1993).

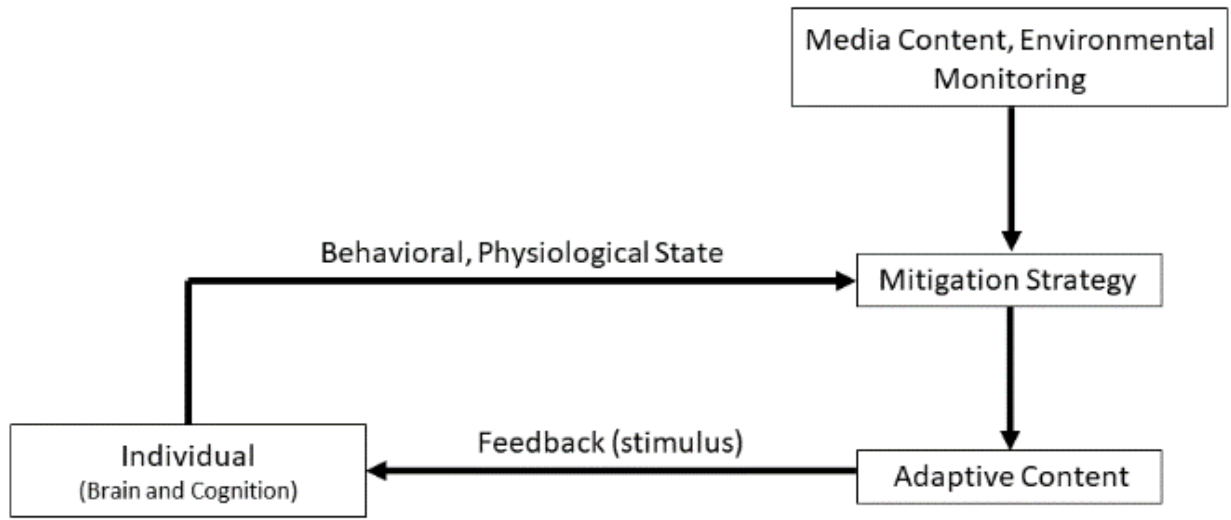

Figure 3. An instance of closed-loop mitigation with generic elements. Briefly, the behavioral and cognitive state of an individual is used to select certain types of media content and/or environmental features using the mitigation strategy. This produces adaptive content in the form of selectively presented content or affordances (Gielo-Perczak & Karwowski, 2003; Good, 2007) that improve overall performance.

**Augmentation and Bio-Psychological Change**

Virtual reality and video games (collectively referred to here as VEs) provides us with a model experimental system for studying the effects of practice, skill transfer, and plasticity-related changes that result from training. Training also allows us to understand augmentation outside the context of real-time mitigation. In the context of VEs, training has been used in the HFE literature for many years as a stand-in for driving and flight simulation (Stedmon & Stone, 2001). Bavelier et.al (2012) review many potential training regimes that are likely to allow skill transfer to real-world tasks. Using this definition, learning is a very broad phenomenon, and encompasses biological processes as broad as development, repetitive performance of novel tasks, musical and athletic training, and action video game playing.

The success of VE in such contexts results from the successful transfer of procedural learning that occurs between the simulation and the real world. The physical effects of training (Adamovich et.al, 2009), and perhaps the transfer of training, allows for augmentation of cognitive and sensory abilities in a controlled context. The broad diversity of training regimes described by Bavelier et.al (2012) allows for a host of performance improvements after a lengthy period of training. There include, but are not limited to: reaction time, attention, and improvements to both top-down and bottom-up attentional processing. In particular, action video games (emphasizing the engagement of hand-eye coordination and speed-accuracy trade-offs) tend to improve both top-down and bottom-up attentional processing (Bavelier et.al, 2012).

**Biological Plasticity and Augmentation**

As transfer effects have been associated with short-term training (Baldwin & Ford, 1988), phenotypic plasticity might also result from longer-term augmentation. In many animals, environmental plasticity is an adaptive change that occurs within the organism's life history (West-Eberhard, 2003). Environmental plasticity is triggered by living under challenging conditions (e.g. draught, food scarcity, cold weather), but are caused by multigenerational natural selection and phylogenetic constraints. While humans cannot regenerate lost limbs, we can learn new behaviors. There may also significant individual variation in human plasticity, particularly as a response to training.

**Example: Visual Adaptation.** In the case of augmentation with VEs, the question is whether changes due to repeated exposure to such environments truly constitutes plasticity. One effect of longer-term VE use (days to weeks) is analogous to the prism adaptation (Chapman et.al, 2010). Prism adaptations occur when the human wears a thick lens or similar optical device for an extended period of time. Prism adaptations may also effect cognitive functions such as spatial cognition and navigation abilities (Glize et.al, 2017). This experience can trigger aftereffects such as generalized visual plasticity (Huxlin, 2008; Sasaki et.al, 2010; Li, 2016), short-term (hours) visual illusions known as optical aftereffects (Thompson & Burr, 2009), and a shifting of the visuospatial frame of objects as represented in the brain. Visual plasticity such as that triggered by VEs also has much broader direct and indirect neuronal effects (Luaute et.al, 2009).

There also seem to be a case that a broader set of physiological effects (often positive) accompany training and extended exposure to VEs. In structured VEs (such as the video game *Super Mario*), structural changes to the brain that include gray matter adaptations have been demonstrated (Kuhn et.al, 2014). However, the question remains as to whether such isolated observations are truly due to training and adaptation, or whether they are due to other factors. Bavelier et.al (2012) suggest that one effect of VEs that allows for greater cognitive transfer is the phenomenon of "learning to learn", in which individuals learn with greater efficacy when they are exposed to a wider variety of information and exemplars. Yet the core question remains: how do we distinguish epiphenomena from predictable and controllable cognitive augmentation?

**Is Augmentation a Meta-phenomenon?**
Before getting into empirical models for inducing and controlling augmentation, we will witness a debate about a concrete instance of HA: using an experimental design to present individuals with a perceptual stimulus, and then measuring some set of effects. This debate revolves around the effects of training in video games (Shatil et.al, 2014; Fisher et.al, 2016; Greenwood & Parasuraman, 2016; Makin, 2016; Mishra et.al, 2016; Stanmore et.al, 2017), and meta-evaluations of these results (Hambrick et.al, 2014, Simons et.al, 2016) that attempt to place such claims in proper context.

Based on the existing literature, cognitive augmentation via VEs (video games) can induce a host of plastic changes. These include but are not limited to the following: expanded attentional capacity (Green and Bavelier, 2004), increased spatial resolution in vision (Green and Bavelier, 2007), reduced multitasking costs that counter the effects of aging (Anguera et.al, 2013), all with effects that persist from minutes to days to months. Contrast this with effects of an intervention such as direct-current stimulation, which has a much smaller effect on mitigating the effects of aging (Nilsson et.al, 2017). In the context of HA, inducing behavioral plasticity and other cognitive effects is an alternative to mitigation strategies such as optimizing attentional load based on the Y-D curve.

In a meta-evaluation of the brain training literature, Simons et.al (2016) found that while brain training interventions improve performance on trained tasks, there is a decreased effect with respect to related tasks. While direct comparisons between studies and outcomes are limited, it seems that generalized effects related to brain training tasks are much less prevalent as compared to task-specific effects. There is a core experimental design that can uncover the desired effects: a three condition experimental design consisting of no test (control), treatment without retest, and treatment with retest. The last condition seems to yield the desired improvements, while the control condition provides a means to evaluate those improvements.

**Augmentation and Experimental Design.** There are also specific types of training that induce cognitive improvements and mitigate limitations to optimal performance. One experimental paradigm that clearly demonstrates cognitive augmentation is Useful Field of View (UFoV) training (Edwards et.al, 2018). In this approach, the idea is to expand the area one can visually survey in a static glance (at a fixed eye and head position). A range of studies have demonstrated two basic results: UFpV decreases with age, but can be expanded as compared to baseline with training. In a related fashion, Deveau et.al (2014) used multisensory percpetual training with diverse stimuli to improve performance in college baseball players in a wide range of statistical categories. This demonstrates a set of conditions for creating a generalized set of effects due to training. These effects not only include transfer to different but structurally similar tasks, but so-called far transfer (Greenwood & Parasuraman, 2016) from one cognitive domain (e.g. attention) to another (e.g. general intelligence).

A second experimental paradigm for clearly demonstrating cognitive augmentation comes from Mishra et.al (2014), and introduces us to Adaptive Distractor Training (ADT). According to this experimental paradigm, the goal of ADT is to select the correct tone in a range of presented tone. The difference in magnitude between the target and distractor tones is then adaptively minimized, and forces a greater degree of auditory discrimination across the training regimen. The resulting learning-induced plasticity can be demonstrated both across the lifespan and between species (Mishra et.al, 2014). This form of augmentation also demonstrates that a mitigation strategy can be designed to overcome decreased tolerance of noise exhibited by the aging brain (deVillers and Merzenich, 2011).

These are not the only types of experimental designs that can be used to induce augmentation. Yet there are a few issues to keep in mind when designing an experimental induction of augmentation. The first is the magnitude of experimental effect. In many cases, the experimental effects will be limited by both the effect size itself and the generalizability of the cognitive effect. Secondly, there are ways to assess the degree of augmentation independent of experimental manipulation and observed effect. This includes modeling techniques beyond the scope of conventional training methods. This set of challenges points us towards different conceptual and analytical models in order to gain a full appreciation of augmentation's effects.

## Inducing and Controlling Augmentation

Given the debate as to whether or not an augmentative effect truly exists, we will not turn to three general approaches to inducing and perhaps even controlling HA. The three techniques presented here are the performance landscape, use of so-called physical media, and extended phenotype modeling.

### Learning Augmentation on a Performance Landscape

Another cognitive function that involves plasticity is learning and memory. The design of systems to augment learning and memory are a bit more complex than the attentional and perceptual cases. Augmenting learning and memory requires an *n*-dimensional construct called a *performance landscape*. Performance landscapes are a class of multidimensional combinatorial phase space (Kauffman, 1993; Gerrits and Marks, 2014) that can be used to represent the process of augmentation in the context of either discrete or dynamic presentation of a media stimulus. We can represent the unfolding of dynamic behavior in this multidimensional space using three dimensions, which allows us to visualize the augmentation process.

Despite this complexity underlying the performance landscape approach, we can identify and target relevant components of the learning and memory process. While it is difficult to translate this model into real-time feedback, this conceptual selectivity can lead to performance improvements in a number of contexts. This requires us to conceive of and describe learning and memory as a dynamical process. According to this view, the learning process results in a dynamically altered performance landscape resulting in a lessening of recall variability and error (Kelso, 1995).

**State Space Approaches.** A more technical way to consider the performance landscape is to embed performance data into a latent two-dimensional state space (Figure 4). State space models (Chen & Brown, 2013) not only provide a quantitative basis for performance landscapes, but also allows us to utilize advanced methods such as first-order control algorithms, gradient descent methods, and stochastic dynamical equations (SDEs). While gradient descent methods enable mitigation of sub-optimal behavior and cognition in naturalistic (continuous) environments, the use of SDEs are particularly promising in terms of modeling learning acquisition, cognitive disruptions, and fluctuations in naturalistic environments. In Figure 5, fluctuations in the learning function can be mapped onto this space, providing us with a performance landscape. The main benefit of using a landscape is that extreme values for both local and global can be identified as components of a large trend. Another benefit to this type of data structure is the ability to use gradient descent methods, particularly for large datasets. Gradient descent uses a mechanism called "hill-climbing" to explore the extrema in the space, and gradient descent algorithms can help in distinguishing between local and global maxima (see Figure 5). Ultimately, this non-real time method might allow augmented individuals to discover new vistas of performance improvements due to training.

Depending on the structure of the underlying task, the performance landscape can be either smooth or rugged. The relative smoothness of the landscape is determined by task complexity. When the environment is completely unstructured, we consider individual points on the landscape as uncorrelated. This results in a highly rugged surface. (Kauffman and Levin, 1987). In a stimulus environment such as a driving simulator or action video game, there is much underlying task structure that allows for correlation amongst many points on the landscape. An example of an uncorrelated landscape would be one that represents a novice user forced to interact with a poorly designed interface.

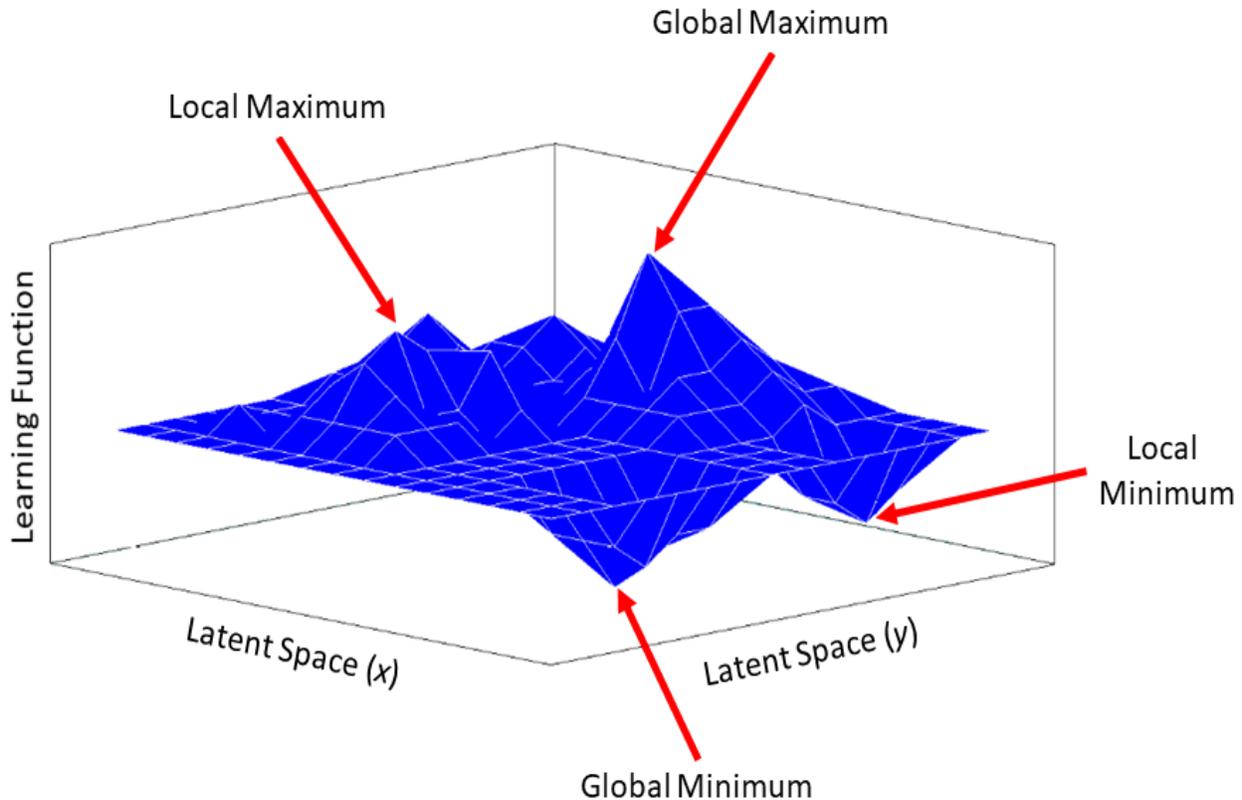

Figure 4. An example of a continuous learning landscape based on pseudo-data. The landscape (blue surface) is defined by a two-dimensional latent space, and a third dimension of depth representing a learning function. The surface also features local and global extrema, in this case representing both positive (maxima) and negative (minima) changes in the learning function.

**Augmentation using Physical Media**

With the miniaturization and portability of communication devices, it makes sense to conceive of media that can produce physical immersion and communication similar to the perceptual properties of VEs. By using so-called physical communication media, we can augment a set of senses such as touch and proprioception in addition to affecting mental models of naïve physics (Gelman and Noles, 2011) and electromechanical flows (Gentner and Gentner, 1982) not typically augmentable using more traditional communication devices. This has been explored a set of experiments (Alicea, 2008b; 2011) that utilized movement against various surfaces and materials to create the illusion of unique and unknowable environments. This would be analogous to interacting with a highly viscous (gelatin-like) atmosphere or non-Newtonian alterations in rotational gravity.

There are two main reasons why this type of training might be interesting to HA practitioners. One is that simulation of such environments might be done to make video games fantasies more immersive and realistic. Training on such environmental analogues might expand the range of sensory representation in people who interact with such technologies, which could contribute to more general performance improvements in sensory and motor abilities. Potential augmentation through these means requires its own model of perceptual and behavioral adaptation. One way this can be done is by considering such experiences as naturally supervised learning (Alicea, 2011). Another means is to consider dynamic processes such as Stochastic Resonance (Alicea, 2009) that rely on interacting sources of noise to produce a cognitive scaffolding (Belland et.al, 2013) for future augmentation.

**Extended Phenotype Modeling**

One quasi-analytical approach to HA involves conceptualizing embodied models of augmentation. To do this, we can take inspiration from fields ranging from Cognitive Science to Psychophysiology, and from Human Factors Engineering and Ethology. Although there is great intellectual diversity featured here, the basic idea is that of a representational space that incorporates biological and environmental features. According to our view, these types of embodied models all describe different aspects of an extended phenotype (Laland, 2004) can be engaged with and modified through various forms of augmentation and training.

**Mobile Infospaces and Reach Manifolds.** The first embodied model is called Mobile Infospaces (Biocca et.al, 2005). Mobile Infospaces are spaces around the torso of an individual augmented with informational displays. This is single-sensory augmentation involving virtual objects located within close spatial proximity to the individual. Since these virtual objects can be both seen and touched, they lie within or become incorporated into the so-called body schema (Mendoza, 2011; Cardinali et.al, 2009a). As individuals interact with objects that are immediately in front of major body parts and within arm's reach, they are likely to incorporate these objects into motor planning and physical action (Graziano et.al, 1994).

A parallel concept from the Human Factors literature involves the approximation of reach manifolds (Yang and Abdel-Malek, 2009). Reach manifolds are spaces around the torso defined by complex surfaces that are reachable by the arms of a human individual. This concept differs from the Mobile Infospace in that reach manifolds are restricted to locations in front of the body where arm reach is optimized. The reach manifold and visual envelope overlap, as human can see farther than a pair of extended arms. Thus as is the case with Mobile Infospaces, there are subtle multisensory interactions which play a role in augmentation (Holmes and Spence, 2004; Spence, 2005).

**Sensory Volumes and the Body Schema.** We can bring both of these technologically-oriented concepts together into a formal model of sensory volumes (Schuster, 2008). Sensory volumes come from the animal behavior (ethology) literature, and describe the continuous space around an individual that is represented by that individual's various senses (Figure 5A). While in animal models sensory volumes typically describe active sensation (fields generated by the organism - for an example see Synder et.al, 2007), in humans they can be used to describe various components of perception and action. For example, a sensory volume of a human would include a visual volume, a haptic/proprioceptive volume, a reach volume, and an auditory volume. Importantly, these volumes overlap to include multisensory spaces. Thus, sensory volumes may also be useful in characterizing changes in spatially-extended sensory capabilities due to augmentation.

To see how augmentation can be described using these quantitative constructs, we can look to the literature describing tool use as an extension of the body schema. Maravita and Iriki (2004) demonstrate this is non-human primates through long-term use of a rake to retrieve food. In Figure 5B, a sensory volume is changed due to long-term interaction with a tool. This demonstrates the the main advantage of extended phenotype models: establishing an explicit set of relationships between physiological mechanisms and embodiment using directly observable behavioral variables (Turner, 2002; Turner, 2004; Schaedelin and Taborsky, 2009; Baccarini et.al, 2014; Cardinali et.al, 2009b; Quallo et.al, 2009).

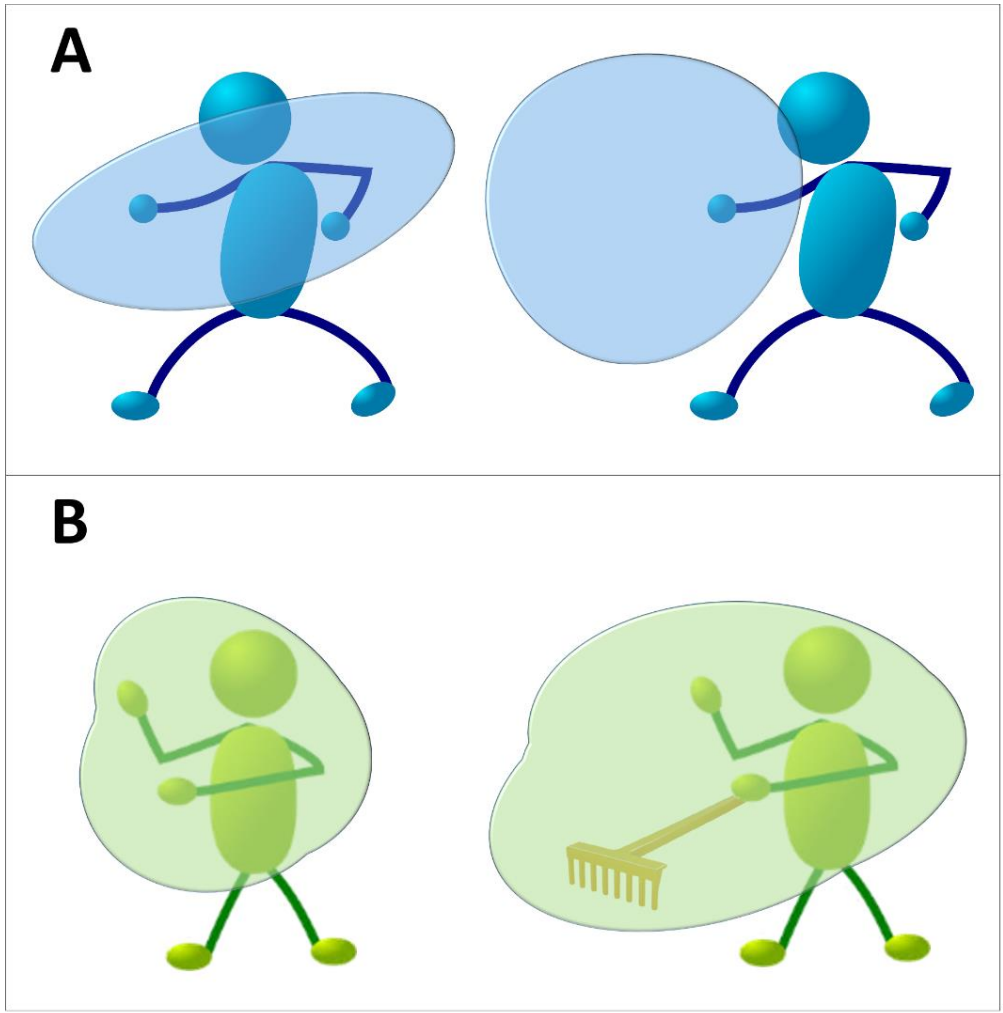

Figure 5. Diagram showing the location and dynamics of sensory volumes. A) diagram showing sensory volumes and their overlap relative to a body. RIGHT: volume for upper body and arm reach, LEFT: reachable volume within visual envelope. B) diagram showing changes in a sensory volume due to tool use. LEFT: area of bodily representation without tool and before tool use, RIGHT: area of bodily representation with tool after tool use.

## Integrative Approaches and Conclusions

To conclude our tour of HA, we can identify avenues for further integrating human biology and technology. This goes beyond the analysis and modeling of data to the biological and physiological systems being subject to augmentation. As the study of biology commonly proceeds in a vertical fashion (from genomes to populations), our approaches to augmented function should ultimately be vertical as well.

**Multilevel Augmentation**

Augmentation is not limited to gains in or attenuation of specific cognitive skills. We must take a very broad view of the issues involved in dynamically changing the perceptual and somatic milieu. Related changes at the cellular/molecular, neurophysiological, behavioral, and social/population levels over time should also play a role in the design of mitigation strategies and the actual causes and effects of augmentation.

**Cellular and Molecular Mechanisms.** The first level of effects to consider are cellular and molecular mechanisms triggered during human performance, and potentially altered through the adaptive process of augmentation. Changes due to focused training and extended practice (such as inflammation) or plastic responses to use (such as tissue regeneration). Many cellular and molecular changes are transient and thus limited to short-term effects. However, transient changes also serve to establish longer-term changes due to training. We can even exploit these changes to benefit the efficacy of a particular mitigation strategy, as cellular plasticity and regulatory changes are associated with changes in synaptogenesis and muscle hypertrophy.

**Neurophysiological Changes.** A second level of effect involves the measurement and real-time monitoring of neurophysiological changes involved in closed-loop regulation (Prinzel et.al, 2000). Measures such as local field potentials, levels of hemodynamic activity, and myogenic potentials provide an indicator of highly complex underlying processes. The development of highly portable and naturalistic techniques such as functional near-infrared (fNIR) imaging (2014) and transcranial direct current stimulation (tDCS) will allow for simultaneous control and measurement of the brain (McKendrick, Parasuraman, & Ayaz, 2015). Networks focusing on anatomic and neurophysiological phenomena (Bullmore & Sporns, 2009) combined with micro-connectomic methods can account for dynamic changes at multiple levels of the brain (Schroter et.al, 2017). Yet even strategically combining methods that measure multiple levels of the nervous system does not provide a greater degree of control with respect to social and environmental phenomena.

**Social and Population-level Effects.** To generalize augmentative systems, we must apply them within and across social groups and biological populations. This explicitly forces us to understand phenomena such as the effects of individual variation (Parasuraman & Jiang, 2012) and social differences in perception (Miller & Saygin, 2013). We must not only understand such phenomena in terms of behavior, but also in terms of how variation becomes manifest at lower levels as well. For example, accounting for social factors such as context can improve the effectiveness of BCIs (Sexton, 2015). Identifying variation across a user population (Szalma, 2009), particularly in terms of baseline characteristics, can also reveal different modes of use and unforeseen limitations of the mitigation strategy.

**Complex mitigation**

There are many cases where the requirements for mitigation are more complex than a simple linear feedback. While attaining optimal attention or motor control are first-order linear problems, many other cognitive processes are both highly complex and involve unknown neural mechanisms.

In such cases, Gershman et.al (2015) propose an approach called computational rationality (CR) as a mitigation strategy. Through a mix of computational architectures and model-free methods, CR allows us to identifying tradeoffs between maximizing utility and minimizing computational costs in complex environments. To ensure minimax performance, we must turn to advances beyond statistical analysis. This includes drawing from work in areas such as computational representation and algorithmic design.

Returning to the idea of $k$-order control, there are a number of issues and concepts relevant to managing the high-dimensional complexity inherent in many cognitive and behavioral systems. In such cases, we might have multiple measurements of cognitive state, in addition to latent state space models of performance. In this type of mitigation, complexity increases exponentially as the number of discrete behavioral states and individual measurements increases. These types of systems differ greatly from the linear control case, and resemble a parallel distributed processing (PDP)-type connectionist models (Mayor et.al, 2014) governed by cybernetic principles.

One of these connections to second-generation cybernetics involves the law of requisite variety. As we might recall, the homeostat as a model of regulatory complexity. Battle's (2014) observed tradeoff between connectivity and stability under complexity provides a means to manage the architecture of large, interconnected HA systems. In complex systems more generally, May (1973) and Stone (2016) have observed a fragility tradeoff. The tradeoff involves growth in the number of connections against the selective loss of connections. While linearly increasing the number of interactions makes each element more robust, the loss of any single element has a disproportionate effect on the entire system.

**Conclusion**

Overall, there are several intellectual traditions contributing to modern forms of HA. One way to view the symbiotic human-machine relationship is to invoke the notion of a cyborg or embodied cybernetic system (Biocca, 1997). Recent advances in mediated environments, computation, and biological measurement is finally allowing us to realize the cyborg model of human-technology interaction. In particular, the functional aspects of embodied cybernetic systems enable processes related to or identified as biological plasticity (Clark, 2007). As an interactive stimulus, the mitigation strategy itself often resembles a form of interactive media. Therefore, another aspect of this heritage comes from Marshall McLuhan, who proposed a metaphoric temperature continuum of media. According to this view, media can be either "hot" or "cool" (1964). "Hot" media are perceptually intense, and only require brief and limited participation. By contrast, "cool" media are less perceptually intense, but also requires longer-term and deeper participation. In the context of mitigation strategies, different media types can serve as a stand-in for different environmental task conditions, and underscore the need for tailored mitigation strategies and sources of behavioral and physiological measurement.

**Abbreviations**

Human Augmentation (HA)
brain-computer interface (BCI)
brain-machine interface (BMI)
virtual environments (VE)

Genetic Algorithm (GA)
Artificial Intelligence (AI)
Electroencephalogram (EEG)
Human Factors Engineering (HFE)
Yerkes-Dodson (Y-D)
Augmented Cognition (AC)
Human-Computer Interaction (HCI)
Intelligence Augmentation (IA)
Useful Field of View (UFoV)
Adaptive Distractor Training (ADT)
Stochastic Dynamical Equations (SDEs)
functional Near-InfraRed (fNIR)
transcranial Direct Current Stimulation (tDCS)
Computational Rationality (CR)
Parallel Distributed Processing (PDP)

Szalma JL. (2009). Individual differences in human–technology interaction: incorporating variation in human characteristics into human factors and ergonomics research and design. *Theoretical Issues in Ergonomics Science*, 10(5), 381-397.

Thompson P & Burr D. (2009). Visual Aftereffects. *Current Biology*, 19(1), R11-R14.

Turner, JS. (2004). Extended Phenotypes and Extended Organisms. *Biology and Philosophy*, 19, 327–352.

Turner, JS. (2002). The Extended Organism: The Physiology of Animal-Built Structures. Harvard University Press, Cambridge, MA.

Vieira M, Lebedev MA, Kunicki C, Wang J & Nicolelis MAL. (2013). A brain-to-brain interface for real-time sharing of sensorimotor information. *Scientific Reports*, 3, 1319.

West-Eberhard, MJ. (2003). Developmental Plasticity and Evolution. Oxford University Press, Oxford, UK.

Wolpaw J & Wolpaw EW. (2012). Brain-Computer Interfaces: Principles and Practice. Oxford University Press, Oxford, UK.

Wolpaw JR, Birbaumer N, McFarland DJ, Pfurtscheller G, & Vaughan TM. (2002). Brain-computer interfaces for communication and control. *Clinical Neurophysiology*, 113(6), 767-791.

Yang J & Abdel-Malek K. (2009). Human Reach Envelope and Zone Differentiation for Ergonomic Design. *Human Factors and Ergonomics in Manufacturing*, 19(1), 15–34.
23